\documentstyle[aps,epsfig,12pt,subfigure]{revtex}
\begin{document}
\draft
\title{Clash of symmetries on the brane}
\author{A. Davidson$^1$, B. F. Toner$^2$,\footnote{Present address:
Department of Physics, California Institute of Technology,
Pasadena, California 91125, U.S.A.} 
R. R. Volkas$^2$ and
K. C. Wali$^3$ }
\address{$^1$ Physics Department, Ben-Gurion University of the Negev\\
Beer-Sheva 84105 Israel\\
$^2$ School of Physics,
The University of Melbourne, Victoria 3010 Australia\\
$^3$ Department of Physics, Syracuse University\\
Syracuse, New York 13244-1130, U.S.A.\\
(davidson@bgumail.bgu.ac.il,\ b.toner@physics.unimelb.edu.au,\ 
r.volkas@physics.unimelb.edu.au,\ wali@physics.syr.edu)}
\maketitle

\begin{abstract}
If our $3+1$-dimensional universe is a brane or domain wall embedded
in a higher dimensional space, then a phenomenon we term the ``clash
of symmetries'' provides a new method of breaking  some continuous
symmetries.  A global $G_{\text{cts}} \otimes G_{\text{discrete}}$
symmetry is  spontaneously broken to $H_{\text{cts}} \otimes
H_{\text{discrete}}$, where the continuous subgroup $H_{\text{cts}}$
can be embedded in several different ways in the  parent group
$G_{\text{cts}}$, and $H_{\text{discrete}} <
G_{\text{discrete}}$. A certain class of topological domain wall solutions
connect two vacua that are invariant under {\it differently
embedded} $H_{\text{cts}}$ subgroups.
There is then enhanced symmetry breakdown to the
intersection of these two subgroups on the domain wall.  This is the
``clash''.  In the brane limit, we obtain a configuration with
$H_{\text{cts}}$ symmetries in the bulk but the smaller intersection
symmetry on the brane itself.  
We illustrate this idea using a
permutation  symmetric three-Higgs-triplet toy model exploiting the
distinct $I-$, $U-$ and $V-$spin U(2) subgroups of U(3).  The three
disconnected portions of the vacuum manifold can be treated
symmetrically through the construction of a three-fold planar domain
wall junction  configuration, with our universe at the nexus.  A
possible connection with $E_6$ is discussed.
\end{abstract}
\vskip2pc

\section{Introduction}

The notion of symmetry lies at the base of modern particle theory, as 
exemplified by the standard model. Some symmetries, such as 
electromagnetic gauge invariance, are manifest: the zero-temperature 
vacuum state and all material systems except for superconductors 
exhibit the symmetry in an explicit fashion. The 
Glashow-Weinberg-Salam SU(2)$_{L}\otimes$U(1)$_{Y}$ electroweak
symmetry, on the other hand, is spontaneously broken: the symmetry of 
the Lagrangian is not shared by the vacuum state. In the standard 
model, the self-interactions of elementary Higgs bosons are  
responsible for making the vacuum asymmetric.

A common opinion is that spontaneous symmetry breaking via 
Higgs bosons is not completely satisfactory, because of the 
proliferation of parameters it brings. In the standard model these 
are predominantly Yukawa coupling constants, while in 
extended theories parameters in the Higgs potential can also 
abound. Hierarchies in {\it a priori} arbitrary parameter values 
can also be seen as troubling. For these reasons, and as an end in 
itself, a search for new ways of breaking symmetries is well motivated.

The purpose of this work is not to do away with elementary Higgs 
fields, but rather to show how they can induce more symmetry 
breakdown than allowed by conventional theory. The scope of the paper 
is to illustrate the basic idea through a non-trivial toy model, 
and then to discuss possible future developments (especially a 
connection with $E_{6}$).

In a conventional theory such as the standard model, the Higgs field 
configuration is assumed to be spatially homogeneous, with a vacuum 
expectation value derived by minimising the Higgs potential. However, 
it is well known that completely stable solitonic configurations 
can also exist if the vacuum manifold has the appropriate 
topology \cite{svr}.
Although such configurations have higher energy than the vacuum state, 
their topological stability allows their use as a background field.  In 
this work, we will use domain wall configurations associated with 
spontaneously broken discrete symmetries.  We will show how 
the symmetry group at the 
centre of a domain wall can be smaller than what you get with a homogeneous 
vacuum configuration, through a phenomenon we term the ``clash of 
symmetries''.  It arises when the symmetry group $H$ of the vacuum manifold 
can be embedded in several ways within the parent group $G$. The enhanced 
symmetry breakdown is caused by the clash of the different 
internal orientations of $H$ within $G$. We will use the $I-$, $U-$ 
and $V-$spin U(2) subgroups of U(3) in our toy model.

There is no observational evidence for a domain wall Higgs 
background of the conventional type in our $3+1$-dimensional 
universe \cite{zeldovich}. To use the clash of symmetries
for realistic model building, we will
therefore ultimately have to identify our universe as a submanifold 
of a higher dimensional space \cite{rs,brane}. 
If the submanifold is infinitely thin in the extra 
dimensions, then it is commonly called a ``brane'', with its complement 
termed the ``bulk''. We will identify our universe with the centre of a 
domain wall configuration dynamically induced by Higgs fields which exist in the 
bulk. The symmetry group at the centre of the wall is then the 
symmetry group of our universe. By taking the appropriate limit, the 
domain wall can be made infinitely thin -- our universe becomes a brane.
Actually, in our toy model the most theoretically appealing configuration will 
be a junction of three semi-infinite walls separated from 
each other by angles of $2\pi/3$. In this case, it seems most 
natural to identify our (toy!) universe with the three-way intersection point, 
the nexus. We will call this 
the ``three-star configuration''.\footnote{Although our toy model
will not incorporate gravity, it is reassuring to note that gravity can be
localised to such a brane junction in the Randall-Sundrum scenario, provided
a single fine-tuning between the cosmological constant and brane
tensions is satisfied~\cite{gravloc}.}

The proposal that we live in a domain wall was made long ago \cite{rs}.
In recent times, the study of branes and/or submanifolds has become a major activity. 
Motivations include string theory, Regge-Teitelbaum gravity and the 
hierarchy problem \cite{brane}. It is interesting that our
motivation to consider brane physics is the independent 
argument presented above. Combining the clash of
symmetries idea with other brane-world ideas may be a
fruitful direction for future work.

Working independently and with a completely different motivation, 
Pogosian and Vachaspati recently 
discovered a class of SU($N$)$\otimes Z_{2}$ Higgs-adjoint kinks featuring the clash of 
symmetries idea~\cite{pv}.  They consider an SU($N$)-adjoint
Higgs field $\Phi$, with Higgs potential 
\begin{equation}
V\left(\Phi\right) = - m^2 \mathrm{tr}\left({\Phi^2}\right) 
+ \gamma \mathrm{tr}\left({\Phi^3}\right) +
\lambda_1 \mathrm{tr}\left({\Phi^4}\right) 
+ \lambda_2 \left[\mathrm{tr}\left({\Phi^2}\right)\right]^2.
\end{equation}
In the absence of the cubic term ($\gamma = 0$), there is a
$Z_{2}$ phase symmetry, $\Phi \leftrightarrow -\Phi$, which
is
outside SU($N$) for odd $N \geq 5$.  For example, in the region of Higgs-parameter
space where 
SU(5) breaks to SU(3)$\otimes$SU(2)$\otimes$U(1),
the authors find domain wall solutions, for which the clash of symmetries results in additional
symmetry breaking to SU(2)$^2\otimes$U(1)$^2$.\footnote{It is interesting to
note that stable wall structures can exist even if a discrete symmetry
is explicitly broken~\cite{dtn}. This can apply for wall configurations having
an everywhere non-vanishing Higgs field, a condition satisfied by the
cases of interest in this paper.}

In this context, quite different from the brane-world
motivation, our paper 
will present another example of this type of kink configuration, within 
a U(3)$\otimes$S$_{3}$ three-Higgs-triplet model.  The S$_3$ symmetry is a
permutation symmetry acting on the three Higgs triplets.  Like
Pogosian and Vachaspati, we impose the discrete symmetry by hand,
by restricting terms that may appear in the Higgs potential.  We,
however, utilise a
permutation rather than phase discrete symmetry because it can be
easily generalised to other groups and other Higgs representations.  This is a
significant difference from the Pogosian-Vachaspati scenario, and for
an SU(3) model seems to be a necessary
ingredient.\footnote{It turns out that the exact
SU(3) analogue of the Pogosian-Vachaspati kink does not exist because
of an accidental SO(8) symmetry.}
The threefold structure of our vacuum manifold will lead us to construct 
the three-star wall-junction configuration mentioned above, an 
exercise that is also of intrinsic technical interest \cite{gravloc,junc}.

The rest of the paper is structured as follows. In the next section, 
we introduce the toy model and discuss its one-dimensional kink 
configurations. Section \ref{2d} is devoted to the three-star 
configuration. We describe a possible connection with $E_{6}$ in Section 
\ref{discussion}. Conclusions and future directions are aired in 
Section \ref{conclusion}. The Appendix establishes that the domain
walls exhibiting the clash of symmetries will be globally stable in
a region of parameter space.

\section{Toy model and its one-dimensional kink solutions}
\label{1d}

\subsection{The Higgs potential and the vacuum manifold}

Consider a model with three Higgs triplets $\Phi_{1,2,3}$ interacting 
through the potential
\begin{eqnarray}
    V & = & - m^{2} ( \Phi_{1}^{\dagger} \Phi_{1} + \Phi_{2}^{\dagger} \Phi_{2}
    + \Phi_{3}^{\dagger} \Phi_{3} ) +
    \kappa ( \Phi_{1}^{\dagger} \Phi_{1} + \Phi_{2}^{\dagger} \Phi_{2}
    + \Phi_{3}^{\dagger} \Phi_{3} )^{2} \nonumber\\ 
    & + &
    \lambda_{1} ( \Phi_{1}^{\dagger} \Phi_{1} \Phi_{2}^{\dagger} 
    \Phi_{2}
    + \Phi_{2}^{\dagger} \Phi_{2}\Phi_{3}^{\dagger} \Phi_{3} +
    \Phi_{3}^{\dagger} \Phi_{3}\Phi_{1}^{\dagger} \Phi_{1} )
    \nonumber\\
    & + &
    \lambda_{2} ( \Phi_{1}^{\dagger} \Phi_{2} \Phi_{2}^{\dagger} 
    \Phi_{1}
    + \Phi_{2}^{\dagger} \Phi_{3}\Phi_{3}^{\dagger} \Phi_{2} +
    \Phi_{3}^{\dagger} \Phi_{1}\Phi_{1}^{\dagger} \Phi_{3} ).
    \label{V}
\end{eqnarray}
The symmetry group of this potential is 
\begin{equation}
    G = G_{\text{cts}} \otimes G_{\text{discrete}} =
    \text{SU}(3) \otimes \text{U}(1)_{1} \otimes \text{U}(1)_{2}
    \otimes \text{U}(1)_{3} \otimes \text{S}_{3}, 
\end{equation}
where the U(1)'s are individual overall phase symmetries for the $\Phi$'s.
The diagonal subgroup of the U(1)'s can be merged with SU(3) to form 
U(3), so $G_{\text{cts}} = \text{U}(3)\otimes\text{U}(1)^{2}$ also.
The role of the discrete permutation symmetry S$_{3}$ is to provide 
topological stability for 
domain wall configurations.  The $\lambda_{2}$ term ensures that the 
continuous symmetry has a common SU(3) for all three multiplets, and the 
sign of $\lambda_{2}$ will cause kinks 
displaying the clash of symmetries (``asymmetric kinks'') to have a different energy 
from those that do not (``symmetric kinks'').
We show below that the asymmetric kink has 
lower energy if $\lambda_{2} > 0$, while the
symmetric kink has lower energy if $\lambda_{2} < 0$.

The U(1) phase symmetries are not germane to our analysis. The 
potential in Eq.~(\ref{V}) was chosen purely for simplicity. By 
including terms such as $\Phi_{1}^{\dagger} \Phi_{2} + \Phi_{2}^{\dagger} \Phi_{3}
+ \Phi_{3}^{\dagger} \Phi_{1} + \text{H.c.}$, the symmetry group can 
be reduced to the more elegant U(3)$\otimes$S$_{3}$.
Inclusion of such terms would change the details of our analysis 
but not its spirit.

To simplify the exposition, 
we set $\kappa = 1$ in Eq.~(\ref{V}) 
by rescaling the field and spacetime
coordinates and measure all
mass-dimension quantities in units of $|m|$, which is equivalent to 
setting $m^{2} = 1$. 

A straightforward analysis shows that there exist three degenerate global minima of the
form
\begin{eqnarray}
\text{Vacuum\ I}:\quad & \langle \Phi_{1}^{\dagger} \Phi_{1} \rangle = \frac{1}{2},\quad
    \langle \Phi_{2}^{\dagger} \Phi_{2} \rangle = \langle \Phi_{3}^{\dagger} 
    \Phi_{3} \rangle = 0,& \label{globalminimumI}\\
\text{Vacuum\ II}:\quad & \langle \Phi_{2}^{\dagger} \Phi_{2} \rangle = \frac{1}{2},\quad
    \langle \Phi_{1}^{\dagger} \Phi_{1} \rangle = \langle \Phi_{3}^{\dagger} 
    \Phi_{3} \rangle = 0,&\label{globalminimumII}\\
\text{Vacuum\ III}:\quad & \langle \Phi_{3}^{\dagger} \Phi_{3} \rangle = \frac{1}{2},\quad
    \langle \Phi_{1}^{\dagger} \Phi_{1} \rangle = \langle \Phi_{2}^{\dagger} 
    \Phi_{2} \rangle = 0,&
    \label{globalminimumIII}
\end{eqnarray}
for the parameter region
\begin{equation}
    \lambda_{1} > 0,\qquad \lambda_{1} + \lambda_{2} > 0.
    \label{parameterregion}
\end{equation}
(Positivity of the potential requires the weaker conditions $\lambda_{1} > -3$ and 
$\lambda_{1} + \lambda_{2} > -3$.) Each global minimum of 
Eqs.~(\ref{globalminimumI}--\ref{globalminimumIII}) 
induces the spontaneous breakdowns
\begin{eqnarray}
\text{S}_{3} & \to & \text{S}_{2} \cong \text{Z}_{2},\nonumber\\
\text{U}(3)\otimes\text{U}(1)^{2} & \to & 
\text{U}(2)\otimes\text{U}(1)^{2}.
\end{eqnarray}
At the level of global vacuum configurations, a $G_{\text{cts}}$ 
transformation
can always be used to bring the nonzero $\langle \Phi_{i} \rangle$
into the form $(1/\sqrt{2},0,0)^{T}$. If this is done, then the 
unbroken U(2) and S$_{2}$ subgroups act on the second and third 
entries of the triplets. The vacuum manifold consists of three 
disconnected pieces labelled I-III in Eqs.~(\ref{globalminimumI}--\ref{globalminimumIII}), 
with each piece being the set of all
$G_{\text{cts}}$ transforms of $(1/\sqrt{2},0,0)^{T}$ for the 
non-vanishing $\langle \Phi_{i} \rangle$.

\subsection{Clash of symmetries}

A kink or one-dimensional domain wall configuration interpolates 
between elements of I and II, or II and III, or I and III, with the 
vacuum states reached at spatial infinity, $z = \pm\infty$. (We
will call $z$ the coordinate perpendicular to the wall. For the
purposes of the following mathematics, it does not matter how many
additional spatial directions exist.) 
For the 
sake of the example, focus on I $\leftrightarrow$ II kinks. Let us
use our SU(3) freedom to set the vacuum I state at $z = -\infty$ to be
\begin{equation}
    \langle \Phi_{1} \rangle = \left( 
         \frac{1}{\sqrt{2}}, 0, 0
        \right)^T,\quad
    \langle \Phi_{2} \rangle = \left( 
        0, 0, 0
        \right)^T,\quad
    \langle \Phi_{3} \rangle = \left( 
        0, 0, 0
        \right)^T.
\end{equation}
The unbroken 
symmetry there is clearly U(2)$_{\text{I}}$, where 
U(2)$_{\text{I},\text{II},\text{III}}$ is 
defined to act on the $(23,31,12)$ entries of the 
triplets.\footnote{In the old days these were called $V$-spin, 
$U$-spin and $I$-spin.}
{\it A priori},  the vacuum II state at $z = +\infty$ can be any element of piece II 
of the vacuum manifold. While all such kinks are in the same topological
class, they are energetically distinguished by
the $\lambda_{2}$ term in $V$. Only the lowest energy member of the class 
is guaranteed topological stability (see below).

The extreme cases are given by the symmetric kink for which
\begin{eqnarray}
   &\Phi_{1}(-\infty) = \left( 
         \frac{1}{\sqrt{2}}, 0, 0
        \right)^T,\qquad
    \Phi_{2}(-\infty) = \left( 
        0, 0, 0
        \right)^T,&\nonumber\\
        &\Phi_{1}(+\infty) = \left( 
        0, 0, 0
        \right)^T,\qquad
    \Phi_{2}(+\infty) = \left( 
         \frac{1}{\sqrt{2}}, 0, 0
        \right)^T,&
\end{eqnarray}
and by the asymmetric kink for which
\begin{eqnarray}
   &\Phi_{1}(-\infty) = \left( 
         \frac{1}{\sqrt{2}}, 0, 0
        \right)^T,\qquad
    \Phi_{2}(-\infty) = \left( 
        0, 0, 0
        \right)^T,&\nonumber\\
        &\Phi_{1}(+\infty) = \left( 
        0, 0, 0
        \right)^T,\qquad
    \Phi_{2}(+\infty) = \left( 
         0,\frac{1}{\sqrt{2}}, 0
        \right)^T,&
\end{eqnarray}
with $\Phi_{3}(z) = 0$ for both cases.

The basic clash of symmetries phenomenon is displayed by the 
asymmetric kink. The symmetric kink has U(2)$_{\text{I}}$ unbroken for all $z$.
For the asymmetric kink, the $z = -\infty$ symmetry is U(2)$_{\text{I}}$ 
while the $z = +\infty$ symmetry is the {\it different} group U(2)$_{\text{II}}$.
At all $z$ in between, the symmetry is reduced to
\begin{equation}
    H_{\text{I}\cap \text{II}} = \text{U}(2)_{\text{I}} \cap \text{U}(2)_{\text{II}} = 
    \text{U}(1)_{\text{III}},
    \label{clash}
\end{equation}
where U(1)$_{\text{III}}$ multiplies the third entry of each $\Phi$ 
by the same phase. [The groups U(1)$_{\text{I,II}}$ are similarly 
defined through cyclic permutations of the subscripts, 
and all three should not be confused with
U(1)$_{1,2,3}$.]
Because the symmetry groups at $z = \pm\infty$ have two additional 
U(1) factors given our simplified Higgs potential, the unbroken 
symmetry for $|z| < \infty$ is actually U(1)$_{\text{III}}$$\otimes$U(1)$^{2}$.
The extra generators are easily determined, and we will not display them.

\subsection{Kink profiles}
\label{sec:ansz}
To derive the kink profiles, one solves for static $z$-dependent
solutions to the Euler-Lagrange equations. Adopting
the ansatz
\begin{equation}
\Phi_{1} = (\phi_{1},0,0)^{T}, \qquad \Phi_{2} =
(0,\phi_{2},0)^{T}, \qquad
\Phi_{3} = (0,0,\phi_{3})^{T}, \label{eq:ansz}
\end{equation}
with $\phi_1$, $\phi_2$ and $\phi_3$ \emph{real}, the equations for the asymmetric kinks are
\begin{equation}
    \phi''_{1} = \phi_{1} [ -1\ + 2\phi_{1}^{2} + (2 + \lambda_{1})
    (\phi_{2}^{2} + \phi_{3}^{2})]\quad \text{and cyclic 
    permutations},
    \label{EulerLagrange1}
\end{equation}
where the prime
denotes differentiation with respect to $z$.  We justify the ansatz of 
Eq.~(\ref{eq:ansz})
in Appendix~\ref{ap:stab}, where we shall see that it is both necessary and
sufficient to obtain stable domain wall configurations.

Returning to our I 
$\leftrightarrow$ II example, we set $\phi_{3} = 0$, and rewrite the 
two remaining equations in terms of 
\begin{equation}
    S \equiv \phi_{1} + \phi_{2},\qquad A \equiv \phi_{1} - \phi_{2}
\end{equation}
to obtain
\begin{eqnarray}
    S'' & = & -S + (1 + \frac{\lambda_{1}}{4}) S^{3} + (1 - 
    \frac{\lambda_{1}}{4}) S A^{2},\label{Seq}\\
    A'' & = & -A + (1 + \frac{\lambda_{1}}{4}) A^{3} + (1 - 
    \frac{\lambda_{1}}{4}) A S^{2}.
    \label{Aeq}
\end{eqnarray}
Notice that the value of $\lambda_{2}$ has no effect on the 
asymmetric kink profile. 

The special parameter point 
$
    \lambda_{1} = 4
$
sees the equations decouple. The solutions with the correct boundary 
conditions are then simply
\begin{eqnarray}
    S(z) & = & \frac{1}{\sqrt{2}},\label{S}\\
    A(z) & = & - \frac{1}{\sqrt{2}} \tanh \frac{z}{\sqrt{2}},
    \label{A}
\end{eqnarray}
or, equivalently,
\begin{eqnarray}
    \phi_{1}(z) & = & \frac{1}{2\sqrt{2}} \left( 1 - \tanh 
    \frac{z}{\sqrt{2}} \right),\label{tanh1} \\
    \phi_{2}(z) & = & \frac{1}{2\sqrt{2}} \left( 1 + \tanh 
    \frac{z}{\sqrt{2}} \right).
    \label{tanh2}
\end{eqnarray}
The hyperbolic tangent function is the archetypal kink profile 
because it is analytically simple. If $\lambda_{1} \neq 4$, then kink 
solutions still exist but can usually be found 
numerically only (there is another
analytic solution describing a stable kink for $\lambda_1 = \infty$ which
we discuss below). A feature of the hyperbolic tangent solution is that
$\phi_{1,2}(z) + \phi_{1,2}(-z) = 1/\sqrt{2}$. 

The brane limit corresponds to the wall being infinitely 
thin. To access it, the mass parameter $|m|$ must be reinstated and 
taken to infinity ($z$ becomes $|m| z$). In this limit
\begin{equation}
    \phi_{1}(|m|z) \to \frac{\Theta(-z)}{\sqrt{2}},\qquad 
    \phi_{2}(|m|z) \to \frac{\Theta(+z)}{\sqrt{2}}, 
\end{equation}
where $\Theta$ is the Heaviside function. The function $(1 + 
\tanh |m| z)/2$ is a ``regularisation'' of $\Theta(z)$.

\begin{figure}[ht]
\begin{center}
\epsfig{file=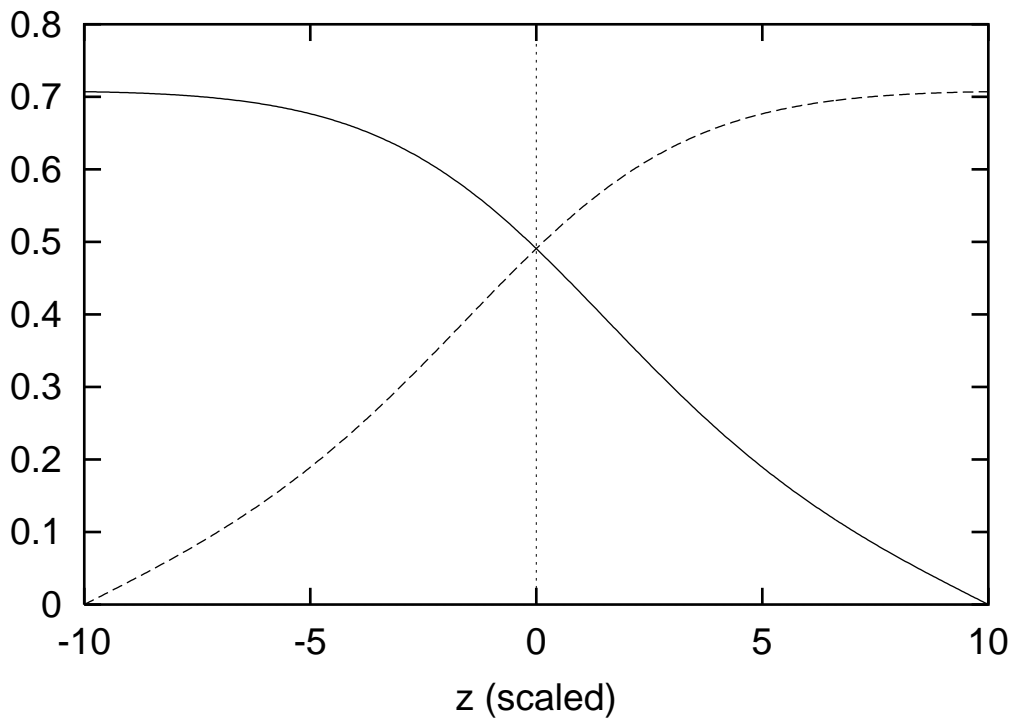,width=14cm}
\vspace{3mm}
\caption{\label{lamp1fig} Asymmetric kink profiles for $\lambda_1 = 0.1$. The solid line
depicts $\phi_1$, and the dashed line $\phi_2$.}
\end{center}
\end{figure}

\begin{figure}[ht]
\begin{center}
\epsfig{file=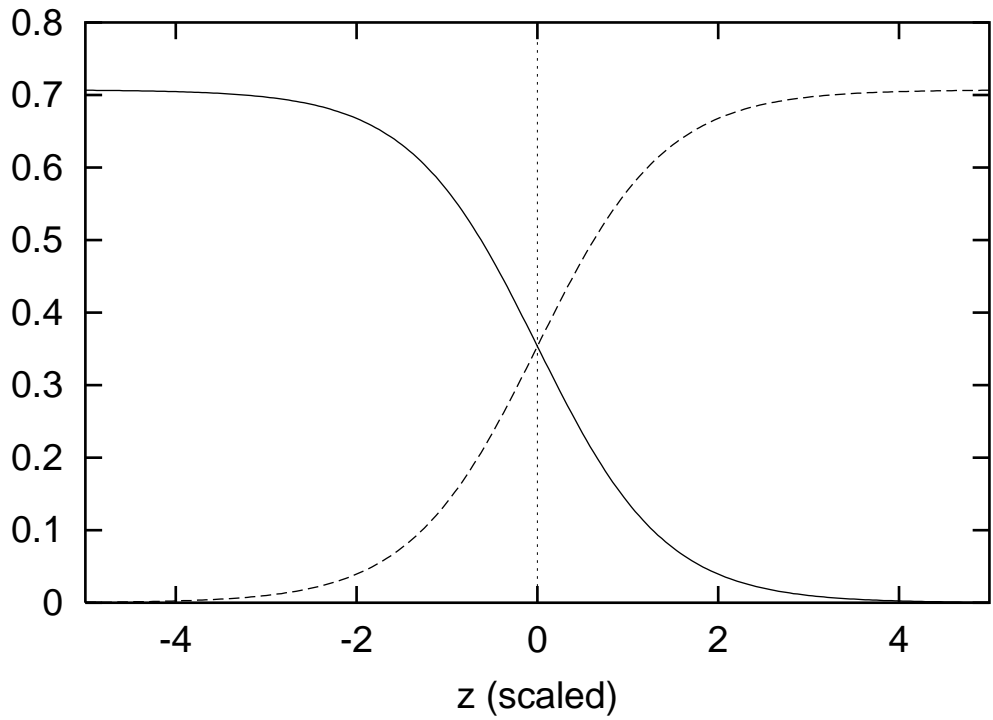,width=14cm}
\caption{\label{lam4fig} As for Fig.\ \ref{lamp1fig} but with $\lambda_1 = 4$.}
\end{center}
\end{figure}

\begin{figure}[ht]
\begin{center}
\epsfig{file=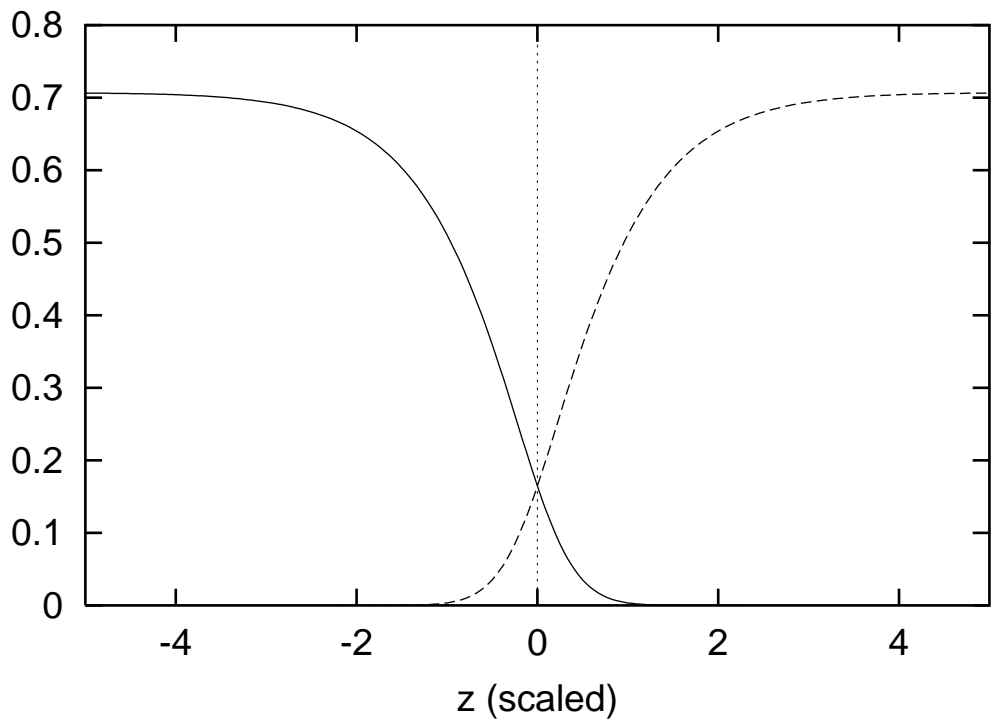,width=14cm}
\caption{\label{lam100fig} As for Fig.\ \ref{lamp1fig} but with $\lambda_1 = 100$.}
\end{center}
\end{figure}

One can push the analysis a little further by using perturbation theory. 
Let $
    \epsilon \equiv \lambda_{1} - 4
$
be a small expansion parameter. Writing
\begin{eqnarray}
    S(z) & = & \frac{1}{\sqrt{2}} +  \epsilon\, \delta\! S(z), \nonumber\\
    A(z) & = & - \frac{1}{\sqrt{2}} \tanh \frac{z}{\sqrt{2}} + 
    \epsilon\, \delta\! A(z),
\end{eqnarray}
substituting in Eqs.\ (\ref{Seq}) and (\ref{Aeq}), equating terms 
$O(\epsilon)$, and solving the resulting equations subject to the
boundary conditions, we find that
\begin{eqnarray}
    \delta\! S(z) & = & \frac{1}{8\sqrt{2}} \left[ \sqrt{2} z \sinh \sqrt{2} 
    z - 2 \cosh \sqrt{2} z\, \ln \left(2\cosh 
    \frac{z}{\sqrt{2}}\right) + 1 \right],\\
    \delta\! A(z) & = & \frac{z}{32} \left( \tanh^{2} 
    \frac{z}{\sqrt{2}} - 1  \right).
\end{eqnarray}
These results give a taste for how the profiles change when one is 
away from the special point but in its neighbourhood. Note that 
$\phi_{1,2}(z) + \phi_{1,2}(-z) = 1/\sqrt{2}$  no longer holds.
Examples of kink solutions are exhibited in Figs.\ \ref{lamp1fig},
 \ref{lam4fig} and \ref{lam100fig}
for $\lambda_1 = $ $0.1$, $4$ and $100$, respectively.
We have confirmed that the perturbative results 
describe the curves well for sufficiently small
$\epsilon$.

\begin{figure}[ht]
\begin{center}
\epsfig{file=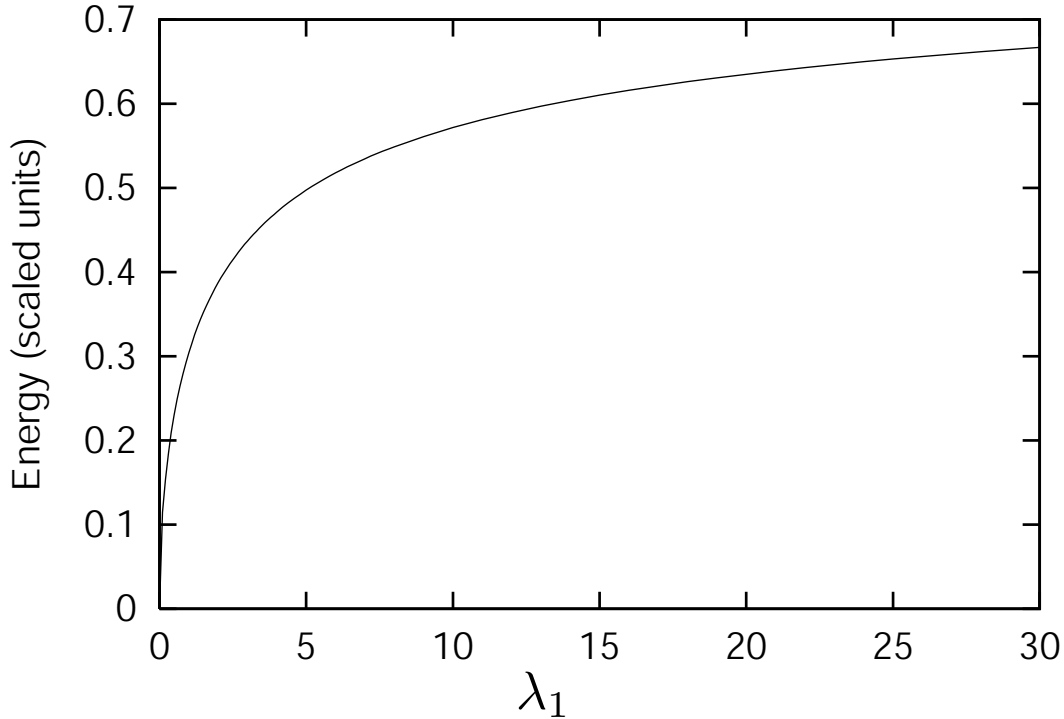, width=14cm}
\vspace{3mm}
\caption{\label{energyfig} Energy per unit area as a function of
  $\lambda_1$ for asymmetric kinks.
For symmetric kinks, replace $\lambda_1$ by $\lambda_1 + \lambda_2$.}
\end{center}
\end{figure}

We next calculate the energy per unit area $\sigma$ of the wall, which is given by 
\begin{equation}
    \sigma = \int_{-\infty}^{+\infty}\, dz\, \left[ (\phi'_{1})^{2} + 
    (\phi'_{2})^{2} + V(\phi_{1},\phi_{2}) + \frac{1}{4} \right],
    \label{energy}
\end{equation}
where the $1/4$ subtracts off the zero point energy. 
At the special parameter point $\lambda_1 = 4$, one obtains
\begin{equation}
    \sigma = \frac{\sqrt{2}}{3}.
\label{energylam4}
\end{equation}
In Appendix~\ref{ap:globstab} we use this result to show that the
solution exhibited in 
Eqs.~(\ref{tanh1}) and (\ref{tanh2}) is globally stable.
For other parameter values, the
energy can be computed numerically, as displayed in Fig.\ \ref{energyfig}.

As mentioned above, there is a second analytic solution valid for all
$\lambda_1$, 
\begin{equation}
\phi_2(z) = \left\{ 
\begin{array}{cc}
     (1/\sqrt{2}) \tanh z/\sqrt{2}&\quad\ \mathrm{for}\ z > 0,\\
  0&\quad\ \mathrm{for}\ z < 0,
\end{array}
\right.
\end{equation}
with $\phi_1(z) = \phi_2(-z)$.  This solution has energy per unit area
$\sigma = 2\sqrt{2}/3$ and though 
perturbatively unstable for all finite $\lambda_1$, it is stable in
the $\lambda_1 \to \infty$ limit. This means that
$2\sqrt{2}/3$ provides an analytic upper
bound on $\sigma$ for all values of $\lambda_1$ and, in particular,
establishes the existence of finite-energy solutions.

It is apparent from Fig.\ \ref{energyfig} that $\sigma$ is
monotonically increasing with $\lambda_1$.  We now establish this
result analytically.
Consider a certain value of
$\lambda_1 > 0$, for which the kink solution is
$\Phi_1 = (\phi_1,0,0)^{T}$, $\Phi_2 = (0,\phi_2,0)^T$.  The $\lambda_1$ term in $V$ is
then simply $\lambda_1 \phi_1^2 \phi_2^2$, which
is non-negative for all $z$.  Therefore if we reduce the value of
$\lambda_1$, the same configuration (which now no longer solves
the Euler-Lagrange equations) has lower energy per unit area.  
But the true solution by definition solves
the Euler-Lagrange equations, so it necessarily has an even lower
energy, energy and action minimisation being equivalent for static
configurations.  

With these results in hand, we may compare the asymmetric kinks with their symmetric 
cousins. To do that, we simply change the ansatz by moving $\phi_{2}$ 
from the second to the first entry in $\Phi_{2}$. The Euler-Lagrange 
equations are the same as Eq.~(\ref{EulerLagrange1}), save for the
substitution
\begin{equation}
    \lambda_{1} \to \lambda_{1} + \lambda_{2}.
\end{equation}
The solutions look very similar, except that the special ``hyperbolic 
tangent point'' is now $\lambda_{1} + \lambda_{2} = 4$.
    
So, for given values of $\lambda_{1}$ and $\lambda_{2}$ obeying
Eq.~(\ref{parameterregion}) 
there are both asymmetric and symmetric 
kink solutions.  Which one is energetically favoured and therefore stable?  We immediately
observe that it depends simply on which of $\lambda_1$ or $\lambda_1 + \lambda_2$
is larger, i.e.\ on the sign of $\lambda_2$.
The clash of symmetries is energetically favoured if
$\lambda_2 > 0$ and energetically disfavoured if $\lambda_2 < 0$.

\section{Planar wall-junction configuration}
\label{2d}

\subsection{Overview and numerical solution}

Our toy model was chosen to produce a vacuum manifold of three 
disconnected pieces I-III as per Eqs.\ (\ref{globalminimumI})-(\ref{globalminimumIII}). 
(The threefold structure is motivated by $E_{6}$, see
Section~\ref{discussion} below.) Each one-dimensional 
kink configuration, however, makes use of only two out of the three 
possibilities. In the context of model building, even if we are only
playing with toys at this stage, it seems more natural to use all three pieces
equally. Perhaps more importantly, clash-induced symmetry breaking will
be enhanced through the presence of all three vacuum types.

\begin{figure}[ht]
\begin{center}
\epsfig{file=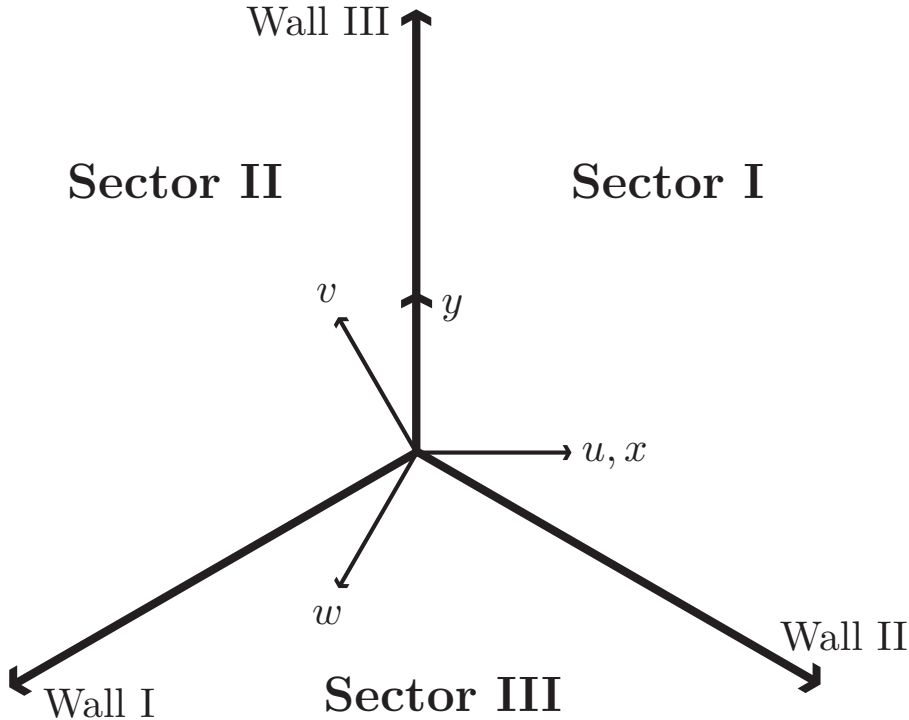,width=12cm}
\caption{\label{3starfig} The three-star domain wall junction configuration.
Each wall makes a $2\pi/3$ angle with the neighbouring walls. In Sector I,
a vacuum I state is reached asymptotically; correspondingly in Sectors II
and III. The two coordinate systems, $(x,y)$ and $(u,v,w)$, are indicated.}
\end{center}
\end{figure}

To that end, we search for a domain wall junction configuration as 
depicted in Fig.\ \ref{3starfig}. Three semi-infinite walls meet at a point, the 
origin or nexus, at angles of $2\pi/3$, dividing the two-dimensional 
plane into three sectors labelled I-III. Let $(r,\theta)$ be the usual 
plane polar coordinates. We impose boundary 
conditions in the obvious way: for a given $\theta$ in sector I, the 
configuration is required to tend to a vacuum I state as $r \to 
\infty$, with corresponding conditions in sectors II and III.
Away from the nexus, and close to a wall, we expect the configuration 
to tend to a one-dimensional kink as a function of the coordinate 
perpendicular to the wall. Let us call this set-up a ``three-star''.
To calculate it, one must solve the equations of motion, this time 
using the two-dimensional Laplacian in place of $d^{2}/dz^{2}$ 
on the left-hand side of Eq.~(\ref{EulerLagrange1}), static and 
$z$-independent solutions being sought.

The brane limit is most conveniently written in terms of the 
Mandelstam-like variables (see Fig.\ \ref{3starfig}),
\begin{equation}
    u = x,\quad v = -\frac{1}{2}x + \frac{\sqrt{3}}{2} y,\quad 
    w = - u - v = -\frac{1}{2}x - \frac{\sqrt{3}}{2} y,
\end{equation}
as
\begin{equation}
    \phi_{1} \to \Theta(u)\Theta(-w),\quad
    \phi_{2} \to \Theta(v)\Theta(-u),\quad
    \phi_{3} \to \Theta(w)\Theta(-v).
    \label{uvwbrane}
\end{equation}
(It is tempting to ``regularise'' this configuration by replacing each 
$\Theta$ with a $(1 + \tanh)/2$. We have checked that this suggestive form captures the 
spirit of the three-star we have produced numerically, but not its detail.) 

There are three different types of three-stars: totally symmetric, 
totally asymmetric, and mixed. The symmetric star has the asymptotic 
vacuum states being cyclic permutations of 
[$\langle \Phi_{1} \rangle = (1/\sqrt{2},0,0)^{T}$, 
$\langle \Phi_{2} \rangle = (0,0,0)^{T}$,
$\langle \Phi_{3} \rangle = (0,0,0)^{T}$]. There is no clash of 
symmetries anywhere for this case: the unbroken symmetry is 
U(2)$_{\text{I}}$ everywhere.

The configuration we want is the totally asymmetric star, defined by 
the vacuum states:
\begin{eqnarray}
    &\text{Sector I:}\quad \langle \Phi_{1} \rangle = 
    (1/\sqrt{2},0,0)^{T},\quad 
    \langle \Phi_{2} \rangle = (0,0,0)^{T},\quad
    \langle \Phi_{3} \rangle = (0,0,0)^{T},&\nonumber\\
    &\text{Sector II:}\quad \langle \Phi_{1} \rangle = 
    (0,0,0)^{T},\quad 
    \langle \Phi_{2} \rangle = (0,1/\sqrt{2},0)^{T},\quad
    \langle \Phi_{3} \rangle = (0,0,0)^{T},&\nonumber\\
    &\text{Sector III:}\quad \langle \Phi_{1} \rangle = 
    (0,0,0)^{T},\quad 
    \langle \Phi_{2} \rangle = (0,0,0)^{T},\quad
    \langle \Phi_{3} \rangle = (0,0,1/\sqrt{2})^{T}.&\nonumber
\end{eqnarray}
Ignoring the superfluous U(1)'s, the clash of symmetries has the 
pattern:
\begin{eqnarray}
    & H_{\text{I}\cap \text{II}} = \text{U}(2)_{\text{I}} \cap \text{U}(2)_{\text{II}} = 
    \text{U}(1)_{\text{III}}\ \text{along wall III},&\nonumber\\
    &H_{\text{II}\cap \text{III}} = \text{U}(2)_{\text{II}} \cap 
    \text{U}(2)_{\text{III}} = 
    \text{U}(1)_{\text{I}}\ \text{along wall I},&\nonumber\\
    &H_{\text{III}\cap \text{I}} = \text{U}(2)_{\text{III}} \cap 
    \text{U}(2)_{\text{I}} = 
    \text{U}(1)_{\text{II}}\ \text{along wall II}.&\nonumber
\end{eqnarray}
At the nexus, the symmetry is completely destroyed:
\begin{equation}
H_{\text{I}\cap \text{II} \cap \text{III}} = 
\text{U}(2)_{\text{I}} \cap \text{U}(2)_{\text{II}} \cap 
\text{U}(2)_{\text{III}} = \{1\}. 
\end{equation}

The totally asymmetric star is energetically 
favoured over the mixed and symmetric stars for the same region of 
parameter space in which the asymmetric kink is favoured over the 
symmetric one. We place our toy $3+1$-dimensional universe at the 
nexus.

\begin{figure}[ht]
\begin{center}
\epsfig{file=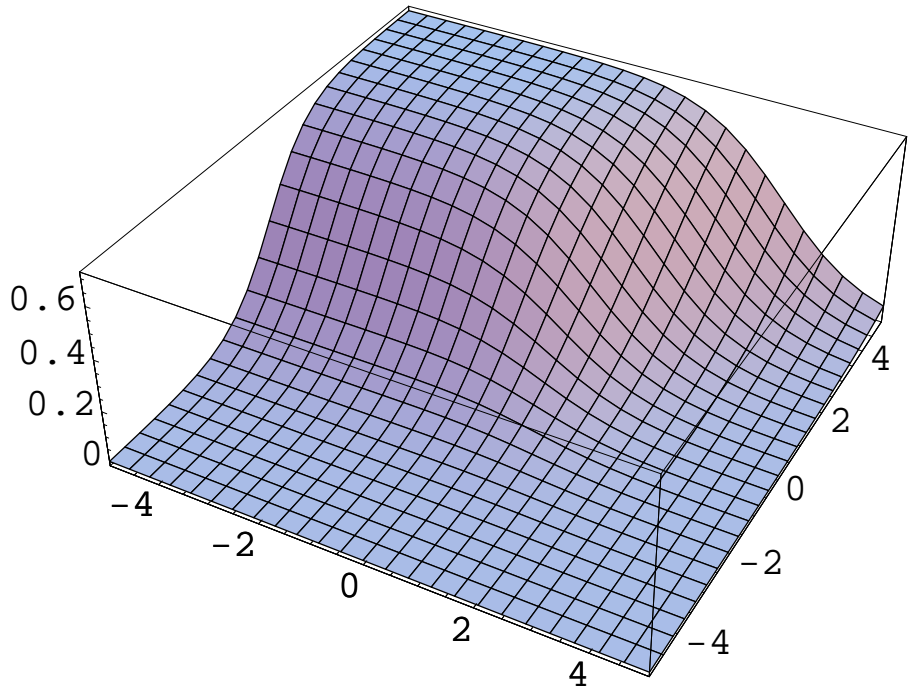,width=12cm}
\caption{\label{p1numericalfig} The $\phi_1$ component of the asymmetric three-star configuration
for $\lambda_1 = 4$ as a function of $x$ and $y$.}
\end{center}
\end{figure}

\begin{figure}[ht]
\begin{center}
\epsfig{file=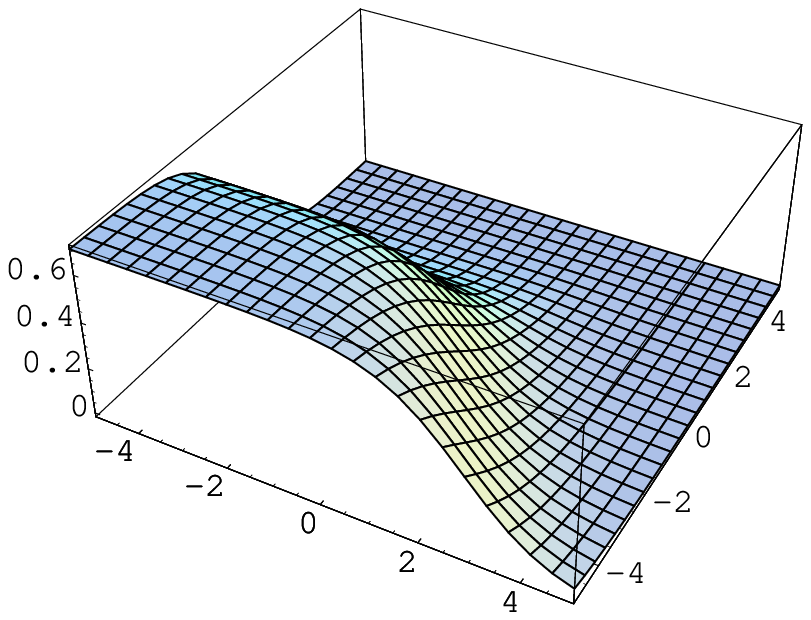,width=12cm}
\caption{\label{p2numericalfig} The $\phi_2$ component of the asymmetric three-star configuration
for $\lambda_1 = 4$ as a function of $x$ and $y$.}
\end{center}
\end{figure}

\begin{figure}[ht]
\begin{center}
\epsfig{file=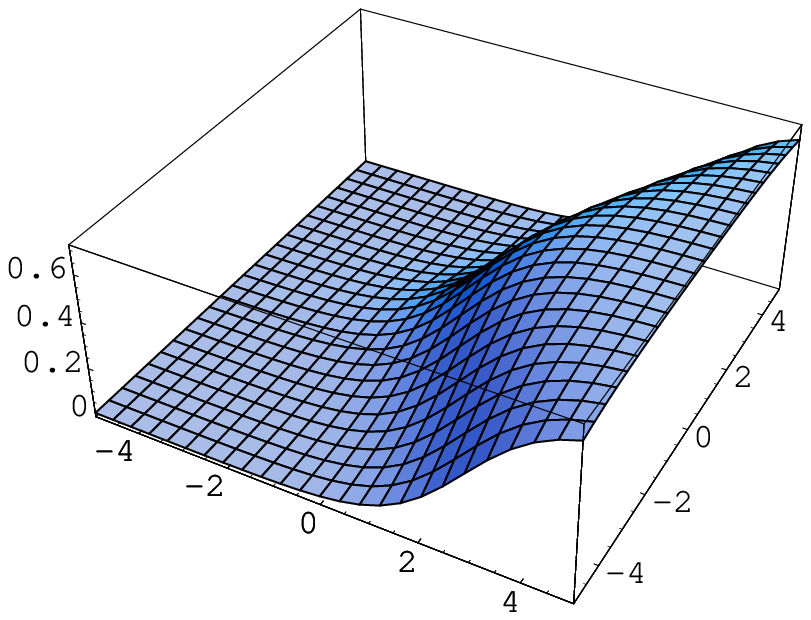,width=12cm}
\caption{\label{p3numericalfig} The $\phi_3$ component of the asymmetric three-star configuration
for $\lambda_1 = 4$ as a function of $x$ and $y$.}
\end{center}
\end{figure}

\begin{figure}[ht]
\begin{center}
\epsfig{file=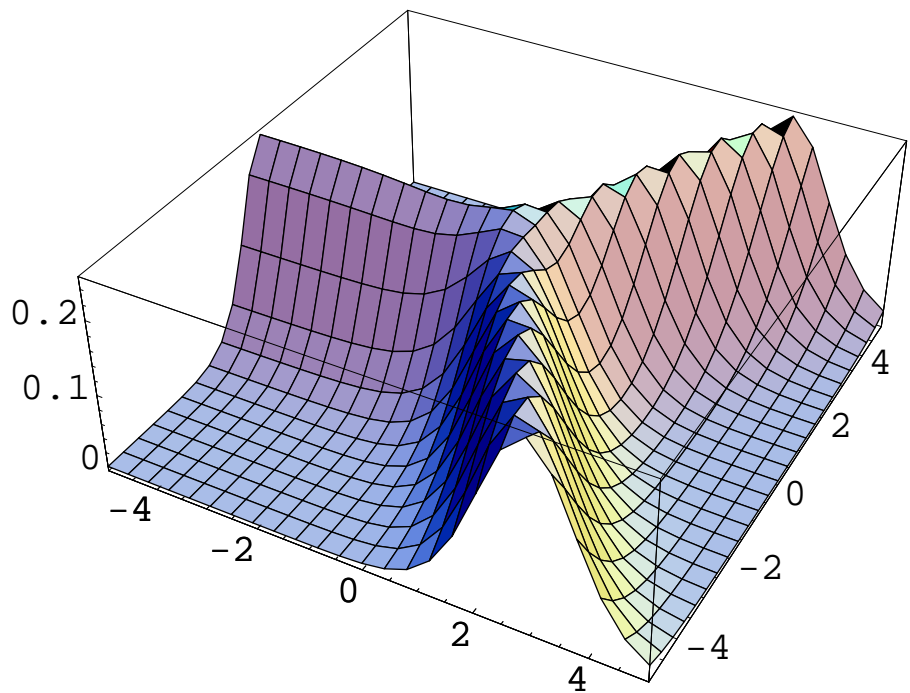,width=12cm}
\caption{\label{energy_numericalfig} The energy density of the asymmetric three-star configuration
for $\lambda_1 = 4$ as a function of $x$ and $y$.}
\end{center}
\end{figure}

Figures \ref{p1numericalfig}, \ref{p2numericalfig}
and \ref{p3numericalfig} show the $\phi_1$, $\phi_2$ and $\phi_3$
components of the numerically computed asymmetric three-star for the 
parameter point $\lambda_{1} = 4$. Figure~\ref{energy_numericalfig}
displays the energy density of the three-star as a function of $x$ and
$y$.  We expect that the three-star,
defined by the $2\pi/3$ angular separation of the walls, is the lowest
energy three-wall junction configuration because it minimises the
total length of the domain walls.  We have checked this numerically: in our simulations,
junctions with unequal angles between the walls relax to
the $2\pi/3$ angular configuration we describe.

\subsection{Partial analytical results}

We now present some analytical results for two different asymptotic 
regimes: off-wall and near-wall, both at large $r$.

Note first, though, that the threefold symmetry of the star implies 
that if
\begin{equation}
    \phi_{1}(r,\theta) = \psi (r,\theta)
\end{equation}
then
\begin{equation}
    \phi_{2}(r,\theta) = \psi (r,\theta - \frac{2\pi}{3}),\qquad
    \phi_{3}(r,\theta) = \psi (r,\theta + \frac{2\pi}{3}).
    \label{threefold}
\end{equation}
Also, the function $\phi_{1}$ must be symmetric under the reflection 
$v \to -v$, $\phi_{2}$ under $w \to -w$, and $\phi_{3}$ under $u \to 
-u$. This is because a sector I vacuum preserves the $\Phi_{2} 
\leftrightarrow \Phi_{3}$ discrete symmetry, and so on.

\subsubsection{Large r, off-wall behaviour}

At large $r$, put
\begin{equation}
    \phi_{1}(r,\theta) = f(\theta) + \epsilon(r,\theta),
    \label{offwallpert}
\end{equation}
where

\begin{equation}
    f(\theta) = \left\{ \begin{array}{cc}
                1/\sqrt{2}&\quad \text{in sector I} \\
	0&\quad \text{elsewhere,} 
       \end{array} \right. 
        \label{f}
\end{equation}
and $\epsilon(r,\theta) \ll 1$. Equation (\ref{threefold}) is then 
used to determine $\phi_{2,3}(r,\theta)$. The perturbative requirement that the 
function $\epsilon(r,\theta)$ be small is met off-wall and at large 
$r$.

Substituting Eq.~(\ref{offwallpert}) into the Euler-Lagrange 
equations, and equating like powers of $\epsilon$ one obtains to 
zeroth order
\begin{equation}
    \frac{1}{r^{2}} \frac{d^{2}f}{d\theta^{2}}(\theta)= f(\theta) \left\{
    -1 + 2 f(\theta)^{2} + (2 + \lambda_{1}) \left[
    f(\theta - \frac{2\pi}{3})^{2} + f(\theta + \frac{2\pi}{3})^{2} 
    \right] \right\}, 
\end{equation}
and to first order
\begin{eqnarray}
    \left[\frac{\partial^{2}}{\partial r^{2}} +
    \frac{1}{r} \frac{\partial}{\partial r} +
    \frac{1}{r^{2}} \frac{\partial^{2}}{\partial \theta^{2}} 
    \right] \epsilon(r,\theta)
    & = &
    \left\{ -1 + 6 f(\theta)^{2} + (2 + \lambda_{1}) 
    \left[ f(\theta - \frac{2\pi}{3})^{2} + f(\theta + 
    \frac{2\pi}{3})^{2} \right] \right\} \epsilon(r,\theta)\nonumber\\
    &\hspace{-4cm}  + & \hspace{-2cm}
    2(2 + \lambda_{1}) f(\theta)
    \left[ f(\theta - \frac{2\pi}{3})\epsilon(r,\theta - \frac{2\pi}{3})
    + f(\theta + \frac{2\pi}{3})\epsilon(r,\theta + \frac{2\pi}{3}) 
    \right].
    \label{1storder1}
\end{eqnarray}
The zeroth order equation is satisfied by the $f$ defined in Eq.~(\ref{f}).

The right-hand side of Eq.~(\ref{1storder1}) must be treated on a 
sector by sector basis. It gives rise to
\begin{equation}
    \left[\frac{\partial^{2}}{\partial r^{2}} +
    \frac{1}{r} \frac{\partial}{\partial r} +
    \frac{1}{r^{2}} \frac{\partial^{2}}{\partial \theta^{2}} 
    \right] \epsilon(r,\theta)
   = \left\{ \begin{array}{cc}
   2 \epsilon(r,\theta) & \quad \text{for}\ \theta\ \text{in sector I},\\
   \frac{\lambda_{1}}{2}\epsilon(r,\theta) & \quad \text{elsewhere,}
    \end{array} \right.
    \label{1storder2}
\end{equation} 
remembering that we must stay off-wall.

To solve Eq.~(\ref{1storder2}), we look for separated variable 
solutions,
\begin{equation}
    \epsilon(r,\theta) = R(r) T(\theta). 
\end{equation}
Substituting in Eq.~(\ref{1storder2}) produces
\begin{eqnarray}
    T''(\theta) & = & - n^{2} T(\theta),\label{T}\\
    R''(r) + \frac{1}{r} R'(r) & = & \left(k^{2} + 
    \frac{n^{2}}{r^{2}}\right) R(r),\label{R}
\end{eqnarray}
where $k^{2} = 2$ for $\theta$ in sector I and $k^{2} = \lambda_{1}/2$ 
otherwise, and we define $k$ to be positive. 
The separation constant is $n^{2}$, and we require it to be
positive. Equation (\ref{T}) is 
solved by $T(\theta) = T_{n}(\theta) \propto \cos (n\theta + \delta)$, where 
periodicity in $\theta$ requires $n$ to be an integer, and $\delta$ is 
determined by requiring symmetry under $v \to -v$ (if $\theta = 0$ is 
the $v = 0$ line bisecting sector I, then $\delta = 0$).

The radial equation (\ref{R}) has general solution
\begin{equation}
    R(r) = c_{1} I_{n}(kr) + c_{2} K_{n}(kr)
\end{equation}
where $c_{{1,2}}$ are constants and $I_{n}$ and $K_{n}$ are
modified Bessel 
functions. The boundary conditions require us to choose
\begin{equation}
    R(r) \propto R_{n}(r) \equiv K_{n}(kr),
\end{equation}
which, for all $n$, has asymptotic behaviour 
\begin{equation}
R_{n}(r) \sim \frac{e^{-kr}}{\sqrt{kr}}.
\label{eq:asymradbeh}
\end{equation}
Writing the general solution to Eq.~(\ref{1storder2}) as a sum over $n$
of $R_{n}(r) T_{n}(\theta)$ with undetermined coefficients, we find 
that
\begin{equation}
    \epsilon(r,\theta) \sim \frac{e^{-kr}}{\sqrt{kr}} F(\theta),
    \label{offwall}
\end{equation}
where $F$ is an undetermined angular function. The nature of $F$ may 
well be different in sectors II and III from sector I, just as 
the inverse decay length $k$ takes different
values if $\lambda_{1} \neq 4$.
Notice that $\lambda_{1} = 4$ is the same special point that 
produces hyperbolic tangent 
kinks. We saw earlier that these kinks have the special property 
$\phi_{1,2}(z) + \phi_{1,2}(-z) = 1/\sqrt{2}$ (taking the I
$\leftrightarrow$ II 
case). The equality of the $k$'s on both sides of the wall is a 
similar special property for the star configuration.

\subsubsection{Large $r$, near-wall behaviour}

We will now explore near-wall behaviour far from the nexus, 
using wall II as our 
example. To begin with, the relevant coordinates are $\rho$, the radial 
distance directly along the wall, and $w$, perpendicular to the 
wall.

We again use a perturbative approach, writing
\begin{eqnarray}
    \phi_{1}(\rho,w) & = & g(w) + \delta(\rho,w),\label{nearwallpert1}\\
    \phi_{2} & \simeq & 0,\\
    \phi_{3}(\rho,w) & = & g(-w) + \delta(\rho,-w),
    \label{nearwallpert3}
\end{eqnarray}
where $\delta$ is small.
The $w$-parity relationship between $\phi_{1}$ and $\phi_{3}$ is 
dictated by the threefold and reflection symmetries of the 
configuration. Although the $\phi_{2}$ field is of order
$\delta(\rho,w)$ along wall II at 
large $\rho$, it enters quadratically into the Euler-Lagrange
equations for $\phi_{1,3}$ so can be set to zero \emph{a priori}.

Note that the regime we explore here is 
physically separated from the off-wall regime probed above, even 
though both lie far from the nexus. If we set $w$ to some finite 
value and take $\rho \to \infty$, then the angular distance from the 
wall goes to zero: $\Delta \theta \sim w/\rho \to 0$. By contrast,
the off-wall region requires finite $\Delta \theta$.

Substitution of Eqs. (\ref{nearwallpert1})-(\ref{nearwallpert3}) into 
the Euler-Lagrange equations yields the zeroth order result
\begin{equation}
    g''(w) = g(w) \left[ -1 + 2 g(w)^{2} + (2 + \lambda_{1}) 
    g(-w)^{2} \right].
    \label{g}
\end{equation}
By symmetry, a similar equation with $w \to -w$ also holds. Defining
\begin{equation}
    S_0(w) = g(w) + g(-w),\qquad A_0(w) = g(w) - g(-w), 
\end{equation}
we recover Eqs.~(\ref{Seq}) and (\ref{Aeq}). This shows very clearly 
that the perpendicular near-wall behaviour far from the origin is 
exactly the appropriate one-dimensional kink. 

The first order analysis depends on the function $g$. To proceed
analytically, we restrict the following to the special $\lambda_1 = 4$
case, so that $g(w) = [1 + \tanh(w/\sqrt{2})]/2\sqrt{2}$.
Setting
\begin{equation}
S \equiv \phi_1 + \phi_3 = S_0(w) + \delta\! S,\quad
A \equiv \phi_1 - \phi_3 = A_0(w) + \delta\! A, 
\end{equation}
substituting in the Euler-Lagrange equations, and equating
terms to first order in $\delta \! S$ and $\delta \! A$ we obtain:
\begin{eqnarray}
\nabla^2 (\delta \! S) &=& \delta \! S(-1 + 6 S_0^2) = 2 \delta \! S, \label{eq:1}\\
\nabla^2 (\delta \! A) &=& \delta \! A(-1 + 6 A_0^2), \label{eq:2}
\end{eqnarray}
where we have also used $S_0(w) = 1/\sqrt{2}$.

We now switch to polar coordinates $(r,\eta)$ where $\tan \eta = w/\rho$
and $r^2 = \rho^2 + w^2$. Searching for separated variable
solutions in these coordinates, we set
\begin{equation}
\delta \! S(r,\eta) = P(r) [W(\eta) + W(-\eta)],\quad 
\delta \! A(r,\eta) = P(r) [W(\eta) - W(-\eta)].
\end{equation}
Substitution in Eq.~(\ref{eq:1}) then immediately yields
\begin{equation}
P''(r) + \frac{1}{r} P'(r) =  (2 + \frac{n^2}{r^2}) P(r),
\label{Peqn}
\end{equation}
where $n^2 > 0$ is the separation constant.
We conclude that
\begin{equation}
P(r) \sim \frac{e^{-\sqrt{2}r}}{\sqrt{\sqrt{2}r}},\quad \forall n
\label{Presult}
\end{equation}
which matches exactly the asymptotic radial behaviour, Eq.~(\ref{eq:asymradbeh}), found earlier
in the off-wall regime. 

Consider now Eq.~(\ref{eq:2}). Using Eq.~(\ref{Peqn}) and the known 
function $g$, 
defining $\Delta(\eta) = W(\eta) - W(-\eta)$, we obtain 
\begin{equation}
  \frac{\Delta''(\eta)}{r^2 \Delta (\eta)} = -3 - \frac{n^2}{r^2} 
+ 3 \tanh^2 \left(\frac{ r \sin \eta}{\sqrt 2} \right).
\end{equation}
For large $r$, the term depending on the separation constant
$n^2$ is suppressed and can be omitted. For small $\eta$,
we can change variables to $w = r \eta$ to get
\begin{equation}
  \frac{\Delta''(w)}{\Delta(w)} \approx -3 + 3 \tanh^2\frac{w}{\sqrt
  2}.
\end{equation}
The solution with the correct antisymmetry in $w$ is then
\begin{equation}
    \Delta(w)  \propto \left[ 3 \tanh \frac{w}{\sqrt{2}}
+ \frac{w}{\sqrt{2}}
    \left( 1 - 3 \tanh^{2} \frac{w}{\sqrt{2}} \right) \right].
\end{equation}

The collection of results above demonstrates that one can make some 
progress in understanding the three-star configuration analytically,
even though an exact analytic solution is at present lacking.

\section{Discussion}
\label{discussion}

Our toy model was chosen not only to be mathematically simple, but also because it
can serve as a prototype for a more realistic theory motivated by E$_6$. While
it is beyond the scope of this paper to explore this connection in detail, we 
would like to comment and speculate on possible future directions. 

The most direct connection is with the maximal SU(3)$^3$ subgroup of E$_6$,
augmented by a discrete Z$_3$ symmetry that rotates the SU(3) factors.
The complex, anomaly-free representation
\begin{equation}
(3,\overline{3},1) \oplus (1,3,\overline{3}) \oplus (\overline{3},1,3),
\end{equation}
which arises from the decomposition of the $27$ of E$_6$, naturally
generalises the three triplet Higgs boson content of the toy model.
As is well known, one generation of quarks and leptons can be placed in
a similar representation. It would be interesting to apply the
clash of symmetries idea in this context, to see what symmetry
breaking patterns can be produced.

In the future pursuit of serious brane model-building, there is no
reason to restrict Higgs potentials to quartic form. If, for instance,
we have a three-star configuration in mind, then the underlying
spacetime is at least $5+1$ dimensional, where renormalisability
requires at most cubic potentials (which are necessarily unbounded
from below and thus presumably unacceptable). The question of
renormalisation should sensibly be deferred until such time as a
connection with a proper theory of quantum
gravity can be made. We recognise that many string-theoretic brane
and brane junction scenarios have already been proposed \cite{string}.

If the full E$_6$ is considered rather than just the SU(3)$^3$
subgroup, then an important issue is domain wall stability.
The Z$_3$ symmetry of the reduced theory, useful for
wall stability, is then presumably embedded within the continuous
E$_6$ symmetry.
According to the general understanding of
defect formation, the breakdown of E$_6$ to, first,
 SU(3)$^3$ $\otimes$ Z$_3$, and then
to some smaller subgroup, will imply the existence of unstable
vortex-wall hybrid structures rather than topologically stable
walls. While it may be possible for the instability timescale
to be very long, a bulk scalar field in the $27$ of E$_6$
offers another natural possibility. Recognising that
singlets arise in the products $27 \otimes \overline{27}$
and $27 \otimes 27 \otimes 27$, we see that a general
Higgs potential will respect a $Z_3$ {\it phase} symmetry
that may not be contained within E$_6$.

The number ``3'' plays a prominent role in the group theory
of E$_6$: there are three SU(3) factors in the maximal
subgroup under discussion, and there are also three ways to embed
electric charge $Q$ within the group. The latter fact has
been remarked on before \cite{Q}, but perhaps it has not received
the attention it deserves. From the perspective of the
subgroup chain E$_6$ $\to$ SO(10)$\otimes$U(1)$''$
$\to$ SU(5)$\otimes$U(1)$'$$\otimes$U(1)$''$, the
three electric charge assignments correspond to
the standard case where $Q$ lies within SU(5), the
flipped SU(5) case where U(1)$'$ is also involved,
and the flipped SO(10) case where U(1)$''$ enters the
definition of $Q$. Now, there is another long-standing
mystery pertaining to the number three: the apparently
superfluous replication of quark-lepton families.
The lack of a compelling explanation despite years
of thought suggests that new approaches should be
seriously considered. We speculate that the three 
$Q$ embeddings, the three-star configuration 
derived from the triply degenerate vacuum structure, and
threefold quark/lepton family replication may be connected.

\section{Conclusion}
\label{conclusion}

Using a model field theory comprising three U(3) Higgs triplets
interacting through a permutation symmetric quartic potential,
we have shown that domain wall and wall-junction solutions exist
displaying the ``clash of symmetries''. This symmetry
breaking mechanism goes beyond standard spontaneous breaking
by exploiting different embeddings of isomorphic
subgroups in the parent group. Our example used the
$I$-, $U$- and $V$-spin U(2) subgroups of U(3). 
We found topologically stable domain wall solutions which asymptote to
vacuum states corresponding to differently embedded
unbroken U(2) subgroups on opposite sides
of the wall. Non-asymptotically, the symmetry
is further broken to the intersection of the
asymptotically unbroken subgroups. This phenomenon
has been previosuly displayed in a different model and with different motivations 
in Ref.~\cite{pv}.
We propose that such a kink-like configuration 
in the thin-wall or brane limit may
exist in a large extra dimension, with our universe
identified with the brane. In that case,
some of the symmetry breaking
in our universe may be due to the clash of symmetries.
Increasing the number of spatial dimensions (notionally) to five,
we numerically constructed a wall-junction three-star 
configuration that exploits
the clash phenomenon to the full,
with the joint or nexus identified with our (toy) universe.
Future work is motivated on several fronts: a possible connection
with E$_6$, a possible connection between the three-star
and threefold family replication, and degree of
freedom localisation to the brane.

\appendix
\section{Justification of the kink ansatz}
\label{ap:stab}
In this appendix, we justify the asymmetric kink ansatz used in Section~\ref{sec:ansz}.
We consider the case $\lambda_2 > 0$ and
show that a globally stable kink must fit this ansatz,
Eq.~(\ref{eq:ansz}).
(Analogous
arguments show that the symmetric kink is globally stable for $\lambda_2 < 0$.)

\subsection{The two-triplet model}

We begin by considering a simpler model with just two triplets and an exchange discrete
symmetry. The Higgs potential is obtained from Eq.~(\ref{V}) by taking $\Phi_3 = 0$.

Consider a general trial solution of the form
\begin{equation}
      \Phi_i(z) = \left(\phi_{i,1}(z), \phi_{i,2}(z),
      \phi_{i,3}(z)\right)^T,
\label{eq:trial1}
\end{equation}
with $i =1, 2, 3$, which satisfies the asymmetric boundary conditions:
\begin{eqnarray}
   &\Phi_{1}(-\infty) = \left( 
\frac{1}{\sqrt{2}}, 0, 0
        \right)^T,\qquad
    \Phi_{2}(-\infty) = \left( 
      0, 0, 0
        \right)^T,&
\nonumber\\
        &\Phi_{1}(+\infty) = \left( 
          0, 0, 0
        \right)^T,\qquad
    \Phi_{2}(+\infty) = \left( 
      0, \frac{1}{\sqrt{2}}, 0
        \right)^T.&
\end{eqnarray}
Define
\begin{equation}
\theta_i = \sqrt{\phi_{i,1}^*\phi_{i,1} + \phi_{i,2}^*\phi_{i,2}+\phi_{i,3}^*\phi_{i,3}},
\end{equation}
and consider a second trial solution of the form 
\begin{eqnarray}
      \Theta_1(z) &=& \left(\theta_1(z), 0,0\right)^T, \nonumber \\
      \Theta_2(z) &=& \left(0,\theta_2(z),0\right)^T.
\label{eq:trial2}
\end{eqnarray}
We will show that the configuration of Eq.~(\ref{eq:trial2}) 
has energy less than or equal to the
energy of the initial trial solution of Eq.~(\ref{eq:trial1}).
It is clear that this configuration satisfies the boundary
conditions.

Consider first the potential energy density $V \left[ \Phi_1, \Phi_2
\right]$.  Clearly $\Phi_1^\dagger
\Phi_1 = \Theta_1^\dagger \Theta_1 $ and
$\Phi_2^\dagger
\Phi_2 = \Theta_2^\dagger \Theta_2 $, so we need only consider the term in the
potential dependent on $\lambda_2$.  But if $\lambda_2 > 0$, then this term is
positive for $\Phi_{1,2}$ of the form of Eq.~(\ref{eq:trial1}) but zero for
the configuration of Eq.~(\ref{eq:trial2}). Thus 
$V \left[ \Theta_1, \Theta_2 \right] \leq V \left[ \Phi_1, \Phi_2
\right]$.

We now turn to the kinetic energy density $T \left[ \Theta_1, \Theta_2
\right]$: 
\begin{eqnarray}
\sum_i {\Theta_i'}^\dagger \Theta_i' & = & \sum_i \left( \theta_i'\right)^2 \nonumber\\
&=& \sum_i \frac{1}{4 \theta_i^2} \left( \phi_{i,1}^* \phi_{i,
1}' + \phi_{i,2}^* \phi_{i,
2}'  + \phi_{i,3}^* \phi_{i,
3}'+ \  \mathrm{c.\ c.}\right)^2 \nonumber \\
&\leq&  \sum_i \frac{1}{\theta_i^2} \left( \left|\phi_{i,1}\right| \left|\phi_{i,
1}'\right| + \left|\phi_{i,2}\right| \left|\phi_{i,
2}'\right|+\left|\phi_{i,3}\right| \left|\phi_{i,
3}'\right|\right)^2 \nonumber \\
&\leq& \frac{\left(\left|\phi_{i,1}\right|^2
+ \left|\phi_{i,2}\right|^2 + \left|\phi_{i,3}\right|^2\right)}{\theta_i^2}\,
\left( \left|\phi_{i,1}'\right|^2
+ \left|\phi_{i,2}'\right|^2 + \left|\phi_{i,3}'\right|^2\right) \nonumber \\
&=& \sum_i {\Phi_i'}^\dagger \Phi_i',
\end{eqnarray}
by the Cauchy-Schwarz inequality.  So $T \left[ \Theta_1, \Theta_2
\right] \leq T \left[ \Phi_1, \Phi_2 \right]$, also.

\subsubsection{The three-triplet Model}

We now add the third triplet to the model.
Staying with $\lambda_2 > 0$ and applying the arguments of the
 previous section, it suffices to
consider a trial solution of the form 
\begin{eqnarray}
      \Phi_1(z) &=& \left[\phi_1(z), 0,0\right]^T, \nonumber \\
      \Phi_2(z) &=& \left[0,\phi_2(z),0\right]^T, \nonumber \\
      \Phi_3(z) &=& \left[0,0,\phi_3(z)\right]^T,
\label{eq:trial3} 
\end{eqnarray}
with $\phi_i$ real.  Note that the boundary conditions on $\phi_3$
require it to vanish asymptotically.
We now establish that this trial solution
has energy greater than or equal to the alternative trial solution,
\begin{eqnarray}
      \Theta_1(z) &=& \left(\sqrt{\left[\phi_1(z)\right]^2 + 
\left[\phi_3(z)\right]^2}, 0,0\right)^T, \nonumber \\
      \Theta_2(z) &=& \left(0,\phi_2(z),0\right)^T, \nonumber \\
      \Theta_3(z) &=& \left(0,0,0\right)^T.
\label{eq:trial4} 
\end{eqnarray}
Observe that $\xi(z) \equiv \sqrt{\left[\phi_1(z)\right]^2 + \left[\phi_3(z)\right]^2}$ obeys
the correct boundary conditions. Now, the kinetic energy density
due to $\xi$ obeys $(\xi')^2 < (\phi'_1)^2 + (\phi'_3)^2$ by the Cauchy-Schwarz inequality.
In the potential energy density function, we need only consider the
$\lambda_1$ term. For trial solution Eq.~(\ref{eq:trial3}) this term
is $\lambda_1( \phi_1^2\phi_2^2 + \phi_2^2 \phi_3^2 + \phi_3^2 \phi_1^2)$,
while for trial solution Eq.~(\ref{eq:trial4}) it is
$\lambda_1( \xi^2 \phi_2^2 ) = \lambda_1( \phi_1^2 \phi_2^2 + \phi_3^2 \phi_2^2 )$.
The latter is obviously smaller than the former.

So globally stable solutions in the two-triplet model are also
globally stable in the full three-triplet model.

\section{Global stability of the $\lambda_1 =4$ analytic solution} 
\label{ap:globstab}
Consider the special parameter point $\lambda_1 = 4$.  We prove global
stability of the analytic solution, Eqs.~(\ref{tanh1}) and~(\ref{tanh2}), by Bogomolnyi's
method~\cite{bogom}.  The energy density $\epsilon\left[\phi_1,
  \phi_2\right]$ of any solution fitting the ansatz given in Eq.~(\ref{eq:ansz}) is 
\begin{eqnarray}
\epsilon\left[\phi_1, \phi_2\right] &=& {\phi_1'}^2 + {\phi_2'}^2 
+ V \left(\phi_{1},\phi_{2}\right) +
\frac{1}{4} \\
&=& \left(\phi_1' + \phi_1^2 +
\phi_2^2 -\frac{1}{2}\right)^2 + \left( \phi_2' + 2\phi_1
\phi_2  \right)^2 
+ \frac{d}{dz} \left( \phi_1 -
\frac{2}{3}\, \phi_1^3 -2 \phi_2^2 \phi_1 \right).
\label{eq:endens}
\end{eqnarray}
The first two terms of this equation are non-negative, so the total
kink energy is 
\begin{equation}
\int_{-\infty}^{\infty} \epsilon\left[\phi_1, \phi_2\right] dz \geq
\int_{-\infty}^{\infty} dz \,\frac{d}{dz} \left({\phi_1}-
\frac{2}{3}\,\phi_1^3 -2 \phi_2^2 \phi_1 \right) 
= \frac{\sqrt 2}{3},
\end{equation}
where we have substituted for the boundary conditions.
Since the analytic solution given in the main text,
Eq.~(\ref{energylam4}) saturates this
lower bound, it is globally stable.

\acknowledgments{BFT was supported by the Commonwealth
of Australia and the University of Melbourne.
RRV is supported by the Australian Research Council
and the University of Melbourne. He would very much like to
thank the Ben-Gurion University of the Negev for the
award of a Dozor Fellowship during which some of this
work was performed. He would also like to thank his
co-authors Aharon Davidson and Kamesh Wali for
great hospitality at their home institutions
while portions of this work were completed. KCW was
supported in part by a grant from NSF, Division of 
International Programs (U.S.-Australia Cooperative Research)
and also in part by a grant from DOE.}

\end{document}